# On the Fermi gas, the Sommerfeld fine structure constant, and the electron-electron scattering in conductors

C. A. M. dos Santos[1], M. S. da Luz[2], F. S. Oliveira[1], and L. M. S Alves[1,3]

[1]Escola de Engenharia de Lorena, Universidade de São Paulo, Lorena - SP, 12602-810, Brazil

[2]Instituto de Ciências Tecnológicas e Exatas, Universidade Federal do Triângulo Mineiro, Uberaba - MG, 38025-180, Brazil

[3]Instituto Federal de Educação, Ciência e Tecnologia Catarinense, Araquari - SC, 89245-000, Brazil

**Abstract:**

Electrical energy is considered as a fundamental parameter for inclusion in Fermi gas theory, in addition to thermal energy. It is argued that electrical energy can move some electrons to above the Fermi Level, providing free charges to carry the electrical current, even at absolute zero temperature. The Drude model, Ohm's law, quantum resistance, and the electrical resistivity due to electron-electron scattering appear naturally as a consequence of the theoretical description, which is based on the quantization of the angular momentum and the Fermi-Dirac distribution, considering total energy as $\epsilon = k_B T + \Phi_0 I$. The electrical and magnetic forces acting on an electron are related to the ratio between the Fermi velocity and the speed of light and show that the electron motion is due to helical paths. Considering the center of mass description for the Bohr atom, it was possible to show that the magnetic force is related to the electrical force as $F_M = (\alpha/\pi)F_E$, which demonstrates that the electrons move in helical paths along the orbit. The helical motion naturally provides for quantization of the magnetic flux, the spin of the electron, and the first correction term of the anomalous magnetic moment. Applying the model to describe the electron-electron scattering allows prediction of the behavior of the electrical resistivity of many metals at low temperatures, which is in excellent agreement with empirical observations.

## 1) Introduction:

Understanding electrical conductivity and how elementary particles interact during electrical transport is one of the goals of solid-state physics. Based on classical mechanics, Drude and Lorentz[1–6] developed the first successful classical theory based on the fundamental idea in which electrical current is carried by free electrons moving randomly in a lattice of atoms. The Drude model was the first to use the idea of a free gas, non-interacting electrons that would move among the positive ion cores. This classical theory provided verification of the Ohm's law and the Wiedemann-Franz law[7,8], but it fails to explain several other experimental observations related to electron-electron, electron-impurity, and electron-phonon interactions. Furthermore, the statistical description of the electrical conductivity cannot take electron-electron scattering into account, which is relevant for describing the electrical transport properties in conventional metals, especially near zero temperature.

Besides the classical Drude model[5,6], the quantum electron theory was developed by Sommerfeld[5,6,9,10], in which quantum laws were applied to describe the electrical conductivity of metals, using the Pauli exclusion principle as a basis[5,6]. According to Sommerfeld[9], conducting electrons move in a constant potential inside the metal and are considered as wave-like particles with quantized energy values.

In addition, the contribution of electron-electron scattering to electrical conductivity in metals has been studied for a long time[11–13]. In the late 1930s, Landau and Pomerantschusk[11], and Baber[12] identified electron-electron scattering and suggested that the collision between electrons of a metal should lead to $T^2$ electrical resistivity dependence. At sufficiently low temperatures, electrical resistivity ($\rho$) follows the simple expression $\rho \sim AT^2$, where $A$ is an intrinsic property of the metal related to the Sommerfeld constant of the electronic heat capacity ($\gamma$) or the Fermi temperature ($T_F$) of the metal[3,11–13]. Nowadays, this behavior has been frequently revisited[14–22] in order to properly describe electron-electron scattering and the electrical behavior of the metals, separating it from the impurities and lattice scattering mechanisms.

On the other hand, quantization of some electrical transport properties, such as quantum resistance[23–25] and Josephson[26–29] effect, have received much theoretical interest[30–32]. The Josephson effect occurs when two superconducting phases are placed in proximity to each other, separated by a thin insulating barrier. The electrical quantization is expressed by the Josephson energy, given by $\epsilon_J \sim \Phi_0 I_C$, where $\Phi_0$ is the magnetic flux quantum and $I_C$ is the superconducting electrical current crossing the junction[28]. Regarding the quantization of the electrical resistance, it was first reported by Klaus von Klitzing in 1980, who observed that the Hall effect becomes quantized when electrons confined in a two-dimensional system are subjected to an external magnetic field[23–25]. In such cases, the electrical quantization is expressed in terms of quantum resistance $R_Q = h/2e^2$, where $h$ is the Planck's constant and $e$ is the elementary charge.

According to Fermi-Dirac statistics, electrons in a metal are distributed in quantized energy levels, where the topmost energy level occupied by electrons, at absolute zero temperature, is called the Fermi energy level, $\epsilon_F$ (ref. [5,6]). All energy levels below $\epsilon_F$ are completely filled by electrons and all energy levels above are completely empty at zero temperature. Heating the metal above absolute zero, the electrons are thermally excited and move up to higher energy levels becoming free to conduct electricity. The distribution of electrons among various energy levels is given by the statistical function called Fermi-Dirac distribution, $f(\epsilon, T)$, which determines the probability of occupation to a given energy level under thermal equilibrium. Based on that, it is well-known that the thermal energy $k_B T$ is the important parameter which controls the number of electrons that move above to the Fermi level, leading to electrical conduction. This leads the following question: How do electrons carry electrical current at absolute zero?

To address this question, we propose that electrical energy is a fundamental parameter required to describe electron conduction at absolute zero. A small increase in electrical energy will move some electrons above the $\epsilon_F$. The associated motion differs drastically from thermal energy, in which electrons move randomly in the metal, since the electrical energy drives free electrons to move in the direction of the field.

Just to give an idea about the electrical energy-electrical current relation, let us consider the first orbit of the Bohr atom[33,34], in which the orbital current is given by

$$I = \frac{e}{\tau_B} = \frac{ev}{\lambda} \tag{1}$$

where $\tau_B$ is the period of one turn around the proton and the orbital velocity is related with the de Broglie wavelength, given by $\lambda = h/mv$, which results in a clear relationship between the electrical energy and the kinetic energy in the orbit, such as

$$\epsilon = \frac{1}{2}mv^2 = \frac{h^2}{2m\lambda^2} = \frac{h}{2e}\frac{ev}{\lambda} = \Phi_0 I \tag{2}$$

where $\Phi_0 = h/2e$ is the magnetic quantum flux, providing a similar energy-current relation described in the Josephson effect[28].

**2) Results and discussion**

*i*) **Fermi-Dirac distribution and electrical energy**

As pointed out in the previous session, in this work we consider the possibility of electrical energy given by $\Phi_0 I$ being included in the Fermi gas theory and evaluate its consequences for the electrical properties.

In the case of the Fermi gas, this implies a new insight in which

$$\epsilon_F = eV_F = k_B T_F = \frac{1}{2}mv_F^2 = \Phi_0 I_F \tag{3}$$

where $\epsilon_F$ is the Fermi energy, $T_F$ is the Fermi temperature, $v_F$ is the Fermi velocity, and $I_F$ is the concept just introduced, hereafter called Fermi current. Just to give an example, Cu, with carrier density $n = 8.49 \times 10^{28}$ m$^{-3}$, has $v_F = 1.57 \times 10^6$ m/s, $v_F = 1.57 \times 10^6$ m/s and $T_F = 8.12 \times 10^4$ K (ref. [5]), resulting in $I_F = 5.42 \times 10^{-4}$ A.

The important question that arises is: what excites electrons above Fermi level if no thermal energy is introduced in the Fermi gas at zero Kelvin? In fact, all the electrons have non-zero energy, but their energy has to do only with the kinetic and potential energy related to the quantized orbits around the nucleus of the atoms. Thus, another source of energy should be introduced in the gas in order to increase the energy of some electrons above the Fermi level at absolute zero temperature to make them free to carry the electrical current. One possibility is introducing electrical energy as mentioned above, although other forms of energy could be added as well, such as magnetic or photonic energy.

In order to see the implications of this possibility, let us start discussing what happens to the Fermi gas statistics, taking into account the electrical energy ($eV = \Phi_0 I$) in the Fermi-Dirac distribution[5,6,35], as follows

$$f(\epsilon, T, I) = \frac{1}{\exp\left(\frac{\epsilon - \epsilon_F}{k_B T + \Phi_0 I}\right) + 1}. \tag{4}$$

Figure 1 displays the density of states as a function of the energy at absolute zero temperature for an electrical energy $\Phi_0 I$, calculated using Equation (4).

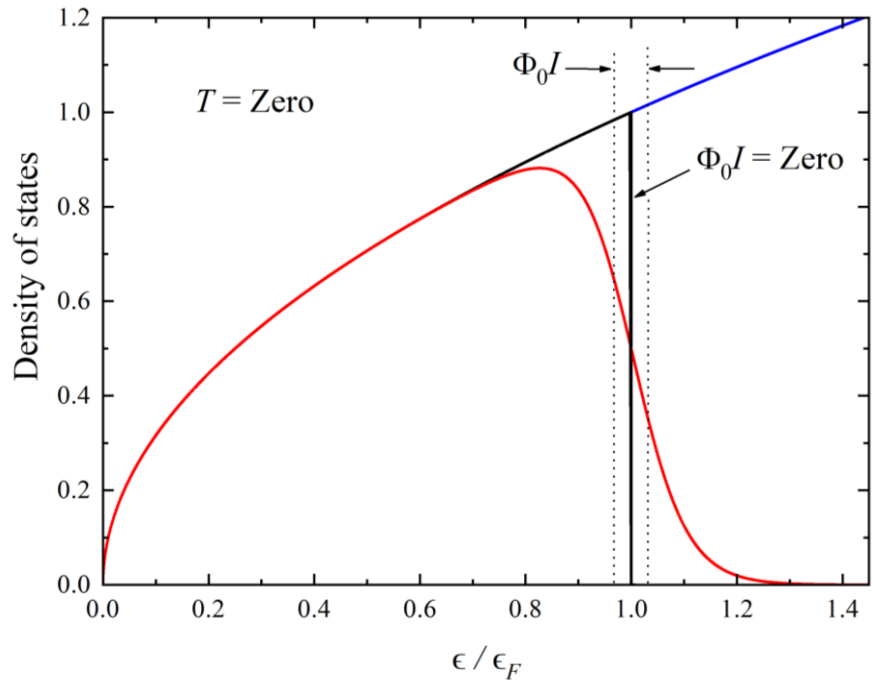

**Figure 1** – Density of states as a function of the energy for a three-dimensional gas (blue line). The black line shows the density of occupied states at absolute zero temperature. At zero temperature electrons will be able to reach energy above the Fermi level ($\epsilon_F$) if an electrical energy ($\Phi_0 I$) is introduced in the Fermi gas. This behavior is represented by the red line, in which the number of electrons excited above the Fermi level ($\epsilon > \epsilon_F$) is proportional to the applied electrical current (the curve was plotted using $\Phi_0 I = 5\% \epsilon_F$). At $T = 0$ and $I = 0$, no electrons are above the Fermi level (black line) and their energies are related only to the kinetic and potential energies due to the quantized orbital motions.

In fact, the distribution is very similar to that considering the thermal energy instead, since Figure 1 was plotted basically exchanging the term $k_B T$ by $\Phi_0 I$ in the Fermi-Dirac distribution (see, for instance, reference 5). However, one must keep in mind that the temperature in Figure 1 is strictly zero and the excitation of the electrons to above the Fermi level is only due to the increase of energy of the Fermi gas by the electrical energy. This has important consequences for understanding the statistical description of the electrical properties, as we will see below.

Let us start looking into the energy of the gas excited by electrical current near the Fermi level.

At $T = 0$ K, one can calculate the energy for the electrons above the Fermi level (see reference 5) as

$$U = \int_{\epsilon_F}^{\infty} (\epsilon - \epsilon_F) D(\epsilon) \, d\epsilon \frac{1}{\exp\left(\frac{\epsilon - \epsilon_F}{\Phi_0 I}\right) + 1} \tag{5}$$

or

$$U(\epsilon_F) = D(\epsilon_F)(\Phi_0 I)^2 \int_0^{\infty} x \, dx \frac{1}{\exp(x) + 1}. \tag{6}$$

after defining $x = (\epsilon - \epsilon_F)/\Phi_0 I$ and considering the limit $\Phi_0 I \ll \epsilon_F$.

Using

$$D(\epsilon_F) = \frac{3N}{2\epsilon_F} \tag{7}$$

and

$$\int_0^{\infty} x \, dx \frac{1}{\exp(x) + 1} = \frac{\pi^2}{12}, \tag{8}$$

one can find

$$U(\epsilon_F) = \frac{\pi^2}{8} \frac{N}{\epsilon_F} (\Phi_0 I)^2, \tag{9}$$

which provides the average of the energy for the electrons near the Fermi level ($\epsilon \gtrsim \epsilon_F$) such as

$$< U(\epsilon_F) > = \frac{U(\epsilon_F)}{N} = \frac{\pi^2}{8} \frac{(\Phi_0 I)^2}{\epsilon_F}. \tag{10}$$

Considering

$$\epsilon_F = \frac{1}{2} m v_F^2 \tag{11}$$

and

$$<U> = \frac{1}{2}m\bar{v}^2, \quad (12)$$

we obtain

$$\bar{v} = \frac{\sqrt{2}}{2}\pi\frac{\Phi_0 I}{mv_F}, \quad (13)$$

where $<U>$ and $\bar{v}$ are, respectively, the kinetic energy and the average velocity of the free electrons right above the Fermi level at zero temperature.

### *ii*) Drude model, Ohm's laws, and quantum resistance

In Figure 2 we define the smallest electrical resistor for a conductor with carrier density $n$, which has the cross section $A = L^2$, whose $L$ is established in the Fermi gas theory[5,6,35] as the side of a cube that contains one electron, calculated by $L = n^{-1/3}$, and the mean free path ($l$), which is related to the Fermi velocity and the relaxation time between collisions ($l = v_F \tau$).

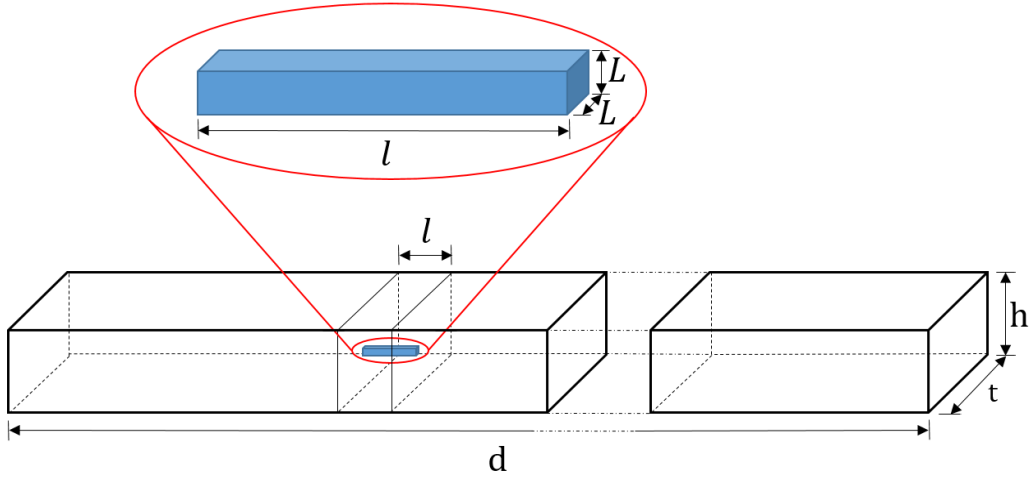

**Figure 2** – Schematic macroscopic sample of length d and cross section ht containing $N$ charges. The magnification represents the quantum electrical resistance taken from any conducting material, which is defined with a cross section $L^2$ and the mean free path $l$. Equations (22) and (23) demonstrate that this resistance is an independent material whose value is given by quantum resistance, $h/2e^2$. Zero temperature electrical resistivity ($\rho_{ee}$) can be found connecting the quantum resistance along with Ohm's law to the quantization of the angular momentum. The combination of parallel and series quantum resistances considering $N$ charges per volume (V = dht) provides the macroscopic electrical resistance ($R$) of the conductor.

Considering the quantum resistance of Figure 2, the first implication of Equation (13) is that the average electrical current predicted by the Drude model[5,6,36], given by

$$nev_d A = I, \quad (14)$$

is recovered after defining the drift velocity as

$$v_d \equiv \frac{\sqrt{2}}{\pi}(3/\pi)^{1/3}\bar{v} = (3/\pi)^{1/3}\frac{\Phi_0 I}{mv_F}, \quad (15)$$

remembering that $A = L^2$ and

$$\lambda = 2L = (3/\pi)^{1/3}\frac{h}{mv_F}, \quad (16)$$

considering Fermi gas theory[5].

This definition allows us to write the microscopically drift velocity as

$$v_d = L\frac{I}{e} \quad (17)$$

where $e/I$ defines the average time related to the electron crossing the distance $L$.

In order to estimate the free number of carriers above the Fermi level due to a small electrical energy ($\Phi_0 I \ll \epsilon_F$) we can write

$$\Delta N = \int_{\epsilon_F}^{\infty} D(\epsilon)d\epsilon \frac{1}{\exp\left(\frac{\epsilon - \epsilon_F}{\Phi_0 I}\right) + 1} \quad (18)$$

providing

$$\Delta N = D(\epsilon_F)\,\Phi_0 I \int_0^{\infty} dx \frac{1}{\exp(x) + 1} = \frac{3N}{2\epsilon_F}\Phi_0 I \ln 2 \quad (19)$$

where $x = (\epsilon - \epsilon_F)/\Phi_0 I$ and

$$\int_0^{\infty} dx \frac{1}{\exp(x) + 1} = \ln 2. \quad (20)$$

After some algebraic calculation taking into account Equations (11), (15), and (19) we can show that

$$\Delta N v_F = 3(\pi/3)^{1/3}\ln 2\, N v_d = 2.112\, N v_d \quad (21)$$

which demonstrates that the number of charges that carry the electrical current near the Fermi level is a very small fraction of the total electrons. For example, a 1 meter long 12-gauge copper wire (cross section = 3.31 x $10^{-6}$ m$^2$, $n$ = 8.49 x $10^{28}$ m$^{-3}$) carrying an electrical current of 1 A has 2.81 x $10^{23}$ electrons available ($v_d$ = 2.22 x $10^{-5}$ m/s), but only 8.37 x $10^{12}$ electrons carry the electrical current ($v_F$ = 1.57 x $10^6$ m/s). The result in Equation (21) provides the equivalence of the Drude model with the Fermi gas theory[5,6,35].

Furthermore, if a small electrical current is applied to the electrical resistance shown in the magnification of Figure 2, the drop in electric potential between two sides of the block over the electrical current must be connected to Equation (3) through

$$\frac{V}{I} = \frac{V_F}{I_F} = \frac{\Phi_0}{e} = \frac{h}{2e^2} = R_Q = 12{,}909 \text{ ohm}. \tag{22}$$

This value represents the quantum electrical resistance, in which the electrical current is carried by a single charge.

Considering the Ohm's law[5,6,8,37] for this resistance, we can write

$$\rho_{ee}\frac{1}{A}l = R_Q \tag{23}$$

where $A = L^2$ and $\rho_{ee}$ is the electrical resistivity of the conductor at $T = 0$, which must be related to the electron-electron scattering, such as will be demonstrated later on. In fact, $\rho_{ee}$ and $L$ are connected in such a way they provide the quantum resistance value ($R_Q$), no matter the conducting material.

Considering a macroscopic sample with $N$ charges ($N$ is comparable to the Avogadro's number) per volume $V = $ dht (see Figure 2 again), one can compute the parallel and series quantum resistances. Noting that ht $= N^{2/3}L^2$ and d $= N^{1/3}l$, the expected electrical resistance of a typical macroscopic sample is

$$R = \rho_{ee}\frac{\text{d}}{\text{ht}}. \tag{24}$$

Note that Ohm's law and the quantum resistance are connected to the quantization of the angular momentum[34] through

$$mv_B^* r_B^* = \hbar = \frac{h}{2\pi}, \tag{25}$$

where $r_B^* = \lambda/2\pi$ and $v_B^*$ are the orbital radius and orbital velocity of the electron in the equivalent Bohr atom, respectively. These are connected to the Fermi velocity

$$v_F = \frac{h}{2\pi m}(3\pi^2 n)^{1/3}, \tag{26}$$

or the wavelength

$$\lambda = 2L = \frac{h}{mv_F}(3/\pi)^{1/3} = \frac{h}{mv_B^*}, \tag{27}$$

providing

$$v_B^* = (\pi/3)^{1/3}v_F = (\pi/3)^{1/3}\frac{l}{\tau} = \frac{l}{\tau^*} = v_F^* \tag{28}$$

where $\lambda$ is the de Broglie wavelength of the electron in the Fermi level of a particular conductor and $\tau^* \equiv (3/\pi)^{1/3}\tau$. This leads to a complete equivalence between $v_B^*$ and $v_F^*$.

Taking Equation (27) along with (23), noting that $v_F = l/\tau$, $L^3 = n^{-1}$, and $A = L^2$, one can show

$$\rho_{ee} = (\pi/3)^{1/3} \frac{m}{ne^2\tau} = \frac{m}{ne^2\tau^*}, \tag{29}$$

which is the well-known expression for the electrical resistivity[8,37]. This demonstrates that the microscopic electrical resistance, schematically shown in the Figure 2, is indeed related to the quantum resistance.

### *iii)* Forces acting on the electron and Sommerfeld fine structure constant

Imagine an electron near the Fermi level moving around the nucleus in the outermost orbit with Fermi velocity ($v_F^* = v_B^*$), schematically shown in Figure 3(*a*). A small increase of electric energy will move it to above the Fermi level. The electron's circular motion will be maintained, but a forward motion due to the force of the electric field will occur, as illustrated in Figure 3(*b*). As a result, an electrical force and a magnetic force act on the electron and its electrical dissipation is given by the Poynting vector, which is associated to the electrical resistance and Joule effect[8,37].

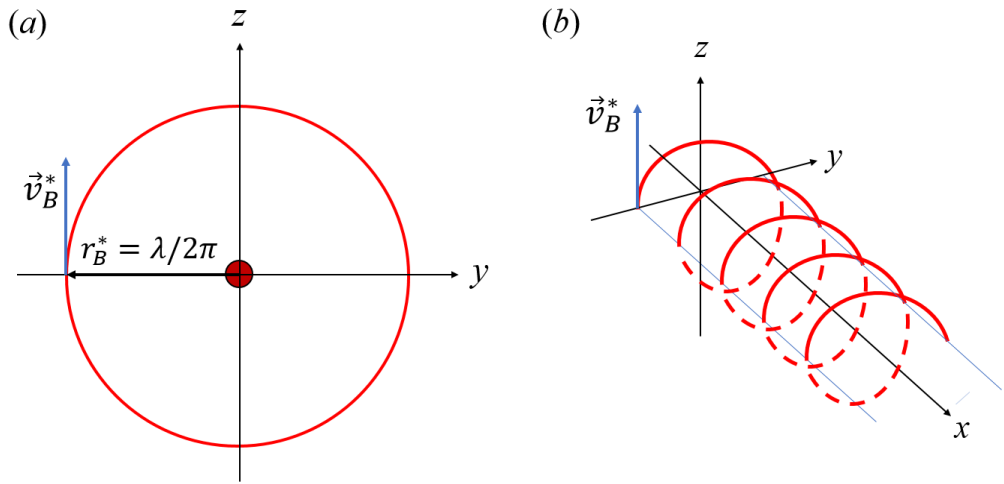

**Figure 3** – (*a*) displays the orbital motion of an electron with equivalent Bohr radius $r_B^* = \lambda/2\pi$ and equivalent Bohr velocity or Fermi velocity $v_B^* = v_F^*$. (*b*) shows the helical motion of the electron as a result of the action of both electrical and magnetic forces.

Now, imagine an electron carrying the electrical current through a quantum resistance discussed in the previous session. The electric force ($F_E$) which makes it to move across the resistance must be given by

$$F_E = \frac{e^2}{4\pi\varepsilon_0(2r_B^*)^2} = \frac{\pi}{16}\frac{e^2}{\varepsilon_0 L^2} \tag{30}$$

where $2r_B^*$ is the diameter of the equivalent Bohr atom or the typical distance between two spheres in the classical collision theory[38], which is related to $L$ or $\lambda$ by

$$2r_B^* = \frac{2L}{\pi} = \frac{\lambda}{\pi}. \qquad (31)$$

As previously mentioned, the dissipation due to the Joule effect in the quantum resistance ($R_Q$) must be related to the electric power ($W_E \propto R_Q I_F^{*2}$) (ref. [8,37]). Thus, one can write the electric power for an electron at the Fermi level interacting with another charge as

$$W_E = F_E v_F^* = \frac{\pi}{16} \frac{e^2}{\varepsilon_0 L^2} v_F^* = C \frac{h}{2e^2} I_F^{*2} \qquad (32)$$

which suggests that an electrical current carried out by one electron crossing the quantum resistance is quantized, where $C$ is constant for a particular conductor. Based on that and using Equation (27), it is easy to show that

$$W_E = \frac{\pi}{16} \frac{e^2}{\varepsilon_0} \frac{8m}{h^2} \epsilon_F^* v_F^* = \pi \frac{\alpha}{\lambda_C} \epsilon_F^* v_F^*, \qquad (33)$$

$$C = \pi \frac{v_B}{v_F^*} = \pi \frac{v_B}{v_B^*}, \qquad (34)$$

and

$$F_E = \pi \frac{v_B}{v_F^{*2}} \frac{h}{2e^2} I_F^{*2}, \qquad (35)$$

after taking into account that

$$I_F^* = \frac{\epsilon_F^*}{\Phi_0} = \frac{2e}{h} \frac{1}{2} m v_F^{*2} = \frac{e v_F^*}{\lambda}, \qquad (36)$$

$$\alpha = \frac{e^2}{2\varepsilon_0 h c}, \qquad (37)$$

and

$$\lambda_B = \frac{\lambda_C}{\alpha} \qquad (38)$$

where $I_F^* = (\pi/3)^{2/3} I_F$, $\epsilon_F^* = (\pi/3)^{2/3} \epsilon_F$, and $v_F^* = (\pi/3)^{1/3} v_F$ (see Equations (3), (27), and (28)), $\lambda_C$ is the Compton wavelength, and $v_B$ and $\lambda_B$ are the orbital velocity and wavelength of the electron in the first orbit of the Bohr atom, respectively.

Using Equations (28), (35), and (37), and remembering that $1/\varepsilon_0 c = \mu_0 c$, it is possible to show that

$$\frac{1}{\pi}\frac{v_B^*}{c}F_E = \frac{\mu_0}{4}\frac{c}{v_F^*}I_F^{*2} = \frac{\mu_0}{4}\frac{l_c}{l}I_F^{*2}, \tag{39}$$

considering

$$\frac{l_c}{l} = \frac{c}{v_F^*}, \tag{40}$$

because

$$\frac{l_c}{c} = \frac{l}{v_F^*} = \tau^* = (3/\pi)^{1/3}\tau, \tag{41}$$

if one assumes $l_c$ as the length which an electron would travel at the speed of light ($l_c = c\tau^*$), instead the mean free path $l$ with the equivalent Fermi velocity ($v_F^*$) during the same time between collisions ($\tau^*$).

Considering the right side of the equation (39) as the magnetic force ($F_M$), which measures the magnetic interaction between the electrical currents by two charges (see, for instance, references 33 and 34) within the distance of the order of the mean free path, one can write

$$F_M = \frac{\mu_0}{4}\frac{l_c}{l}I_F^{*2}, \tag{42}$$

providing

$$F_M = \frac{1}{\pi}\beta F_E = \frac{1}{\pi}\frac{v_B^*}{c}F_E, \tag{43}$$

where

$$\beta \equiv \frac{v_B^*}{c} = \frac{v_F^*}{c}. \tag{44}$$

It is interesting that if $v_B^* = v_B$, the previous equation becomes

$$F_M = \frac{\alpha}{\pi}F_E, \tag{45}$$

which shows that the magnetic force is related to the Sommerfeld fine structure constant[39–43] and is three orders of magnitude smaller than the electrical force for the first orbit of the Bohr atom.

Equation (43) seems to be a fundamental equation and confirms the hypothesis that the two forces act on the electron, one related to the electrical force, which could be either non-dissipative (centripetal) or dissipative, and another related to the magnetic interaction, making the electron move in a helical motion instead of in a straightforward way.

In addition, if one considers the electrical and magnetic fluxes due to the electrical ($F_E$) and magnetic forces ($F_M$), respectively, crossing a closed surface area given by $A = 4\pi(2r_B^*)^2$, it is easy to show from Equations (30), (39), and (42) that

$$EA = \frac{F_E}{e} 4\pi(2r_B^*)^2 = \frac{e}{\varepsilon_0} = \Phi_E \tag{46}$$

and

$$BA = \frac{F_M}{ev_B^*} 4\pi(2r_B^*)^2 = \frac{\mu_0}{\pi} ce = \frac{4\alpha}{\pi} \frac{h}{2e} = \frac{4\alpha}{\pi} \Phi_B . \tag{47}$$

After using Equation (43), we can demonstrate that

$$\Phi_E = 4\alpha c \Phi_B , \tag{48}$$

which provides a direct connection between the electrical and magnetic fluxes with the definition of the Sommerfeld fine structure constant. Furthermore, Equation (48) is the same equation found recently for the energy distribution of an electromagnetic radiation, using a different approach[44].

Now, the questions that arise are: What is the physical meaning of the length $l_c$ and what does it have to do with the electron motion? To address these questions, we will look at the Bohr atom.

### *iv) Helical motion and consequences to the Bohr atom*

In order to have a deeper insight about the $l_c$ length and the helical motion, let us look at the Bohr atom carefully.

Remembering that $v_B/c = \alpha$ for the first orbit of the Bohr atom and supposing that $\lambda_B/l_B$ has to do with the ratio between the Bohr radius ($r_B = \lambda_B/2\pi$) and a free electron traveling at the speed of light that must be proportional to $\alpha$ as well, then $l_B = \lambda_B/\alpha$ or $l_B = \lambda_C/\alpha^2$, since $\lambda_C/\lambda_B = \alpha$ (ref. [39–42]).

As such, electrical and magnetic forces act on the electron but in a way the Poynting[8, 37] vector produces the conservation of the energy, since there is no dissipation in the orbits of the electrons in atoms. In terms of dissipation this is very different from the electrons' motion above the Fermi level described in the previous session.

After spending a long time considering this picture, the only way we have found to take into account both forces simultaneously was to suppose that the electron travels in helical paths around the orbit of the Bohr atom. Figure 4(*a*) displays a schematic view of this helical motion. It considers N = 18,778.865 (= $1/\alpha^2$) spirals per turn of the first orbit in the Bohr atom, leading to a type of toroidal solenoid[8,37].

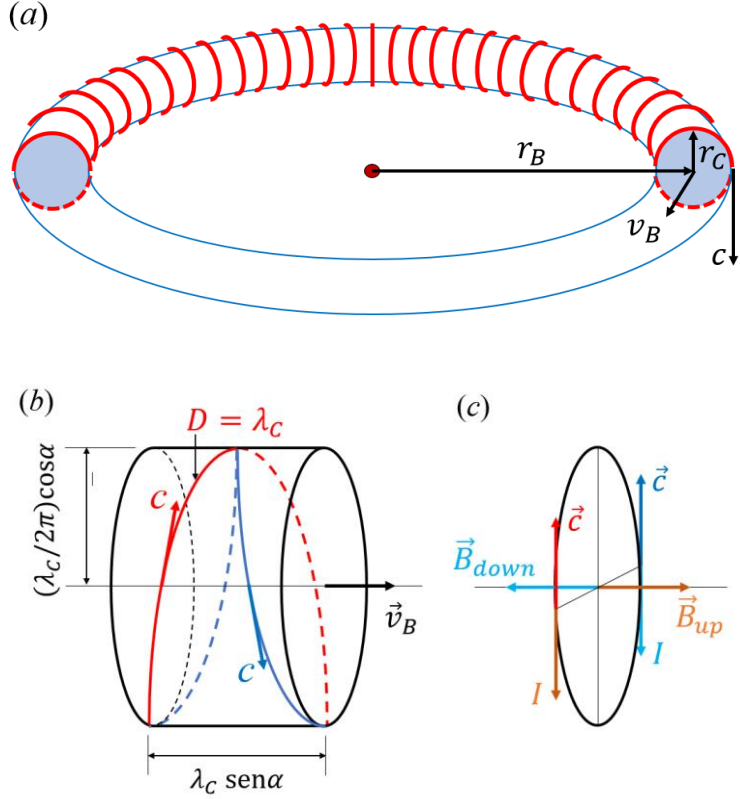

**Figure 4** – (*a*) Helical motion of the electron leading to a toroidal solenoid with 18,778.865 (N = $1/\alpha^2$) spirals. In such a case, the fraction of the $\lambda_B$ orbit with $\lambda_C$ length has 137.036 spirals (= $1/\alpha$). (*b*) Path of one spiral with a radius $r_C \approx \lambda_C/2\pi$ and spacing between adjacent spirals $p = \lambda_C \operatorname{sen} \alpha$ (for viewing purposes, the drawing is enlarged along $v_B$ direction). Red and blue arrows indicate the velocity of the electron which is equal to the speed of light. The projection of $c$ onto the x-axis provides the Bohr velocity of the first orbit, *i. e.* $v_B = c \operatorname{sen} \alpha \approx c\alpha$ (black arrow). (*c*) Since $\lambda_C \operatorname{sen} \alpha \ll (\lambda_C/2\pi) \cos \alpha$, the electrical current in the spiral ($I = ec/\lambda_C$) produces a magnetic field given by Equation (59). The electrical current, the magnetic field, and the cross section of the spiral naturally provide the quantization of the magnetic flux and the electron spin given by equations (61) and (62), respectively. The forwarded helical motions for two electrons in clockwise and counterclockwise directions lead to spin up ($\vec{B}_{up}$) or spin down ($\vec{B}_{down}$), respectively.

In fact, the number of spirals in the solenoid has to do with the ratio $l_B/\lambda_B = 1/\alpha$ or $l_B/\lambda_C = 1/\alpha^2$. Thus, considering $\alpha = 1/137.03599911$ (ref. [43]) and $\lambda_C = 2.462 \times 10^{-12}$ m, one can obtain $l_B = 4.556 \times 10^{-8}$ m for the first orbit of the Bohr atom.

It is interesting that the distance of one spiral ($D$), shown in Figure 4(*b*), can be calculated using

$$D^2 = 4\pi^2 r_C^2 + p^2 \qquad (49)$$

where $r_C = \lambda_C/2\pi \cos \alpha$ is the radius and $p = \lambda_C \operatorname{sen} \alpha$ is the spacing between adjacent spirals, which is very close to the classical electron radius[34], since $\operatorname{sen} \alpha \approx \alpha$. Furthermore, independently of the $\alpha$ angle, the shape of the spiral provides $D = \lambda_C$, which suggests that this must be a universal behavior of electrons in atomic orbits. Furthermore, considering the $l_B = N\lambda_C$ it is easy to show that

$$R_\infty = \frac{\alpha^2}{2\lambda_C} = \frac{1}{2l_B}, \tag{50}$$

where $R_\infty = 8\varepsilon_0^2 h^3 c / m e^4$ is the Rydberg constant for the infinite nuclear mass[45], which is clearly related to the length of the $N$ spirals, suggesting again that the number of spirals is a constant for the orbits of the electrons in the atoms.

Right after calculating both the number of spirals $N$ and the distance $D$, we realized that the velocity of the electron in the first orbit of the Bohr atom is related to the projection onto the orbital direction due to the helical motion of the electron running at speed of light (see red, blue, and black arrows in Figure 4(b)). This suggests that electrons do not travel directly along the orbits but in helical paths at speed of light under a spiral radius given by $r_C \approx \lambda_C/2\pi$, as shown in Figure 4, which seems to agree with Dirac's predictions for the motion of electrons (see, for instance, references 46 and 47).

This description confirms the proportionality relationship between $F_M$ and $F_E$ described in the previous session, as can be seen below. Let us look carefully at the Bohr atom using the center of mass idea, taking into account the motion of both electron and proton, as displayed in Figure 5.

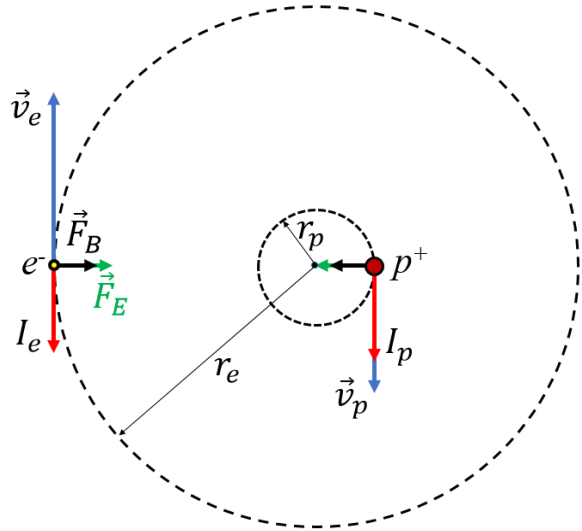

**Figure 5** – Bohr atom considering the center of mass. $v_e$ and $v_p$, $r_e$ and $r_p$, and $I_e$ and $I_p$ are the orbital velocities, the orbital radius, and electrical currents of the electron ($e^-$) and proton ($p^+$), respectively. $F_E$ and $F_B$ represent the electrical and magnetic forces due to the interaction of the charges and electrical currents of the electron and proton. Both forces are centripetal as predicted by Equations (56) and (57).

Using the well-known description of the center of mass for the Bohr atom, the following equations can be found[34]:

$$\frac{v_e}{r_e} = \frac{v_p}{r_p} = \frac{e^2}{2\varepsilon_0 h} \frac{1}{(r_e + r_p)}, \tag{51}$$

$$I_e = e \frac{v_e}{2\pi r_e}, \tag{52}$$

$$I_p = e \frac{v_p}{2\pi r_p}, \tag{53}$$

$$I_T = e \frac{(v_e + v_p)}{2\pi (r_e + r_p)}, \tag{54}$$

$$I_e = I_p = I_T, \tag{55}$$

and

$$\frac{v_e + v_p}{c} = \alpha = \frac{e^2}{2\varepsilon_0 hc} \tag{56}$$

where $v_e$ and $v_p$, $r_e$ and $r_p$, and $I_e$ and $I_p$ are the orbital velocities, the orbital radius, and the electrical currents of the electron and proton, respectively.

A careful inspection of the electrical currents (see Figure 5 again) shows that a magnetic force must exist in the same direction as the electrical force, similarly to two parallel wires carrying the electrical currents $I_e$ and $I_p$ (ref. [8,37]). This magnetic force per length unity ($l_B$) must be given by

$$\frac{F_M}{l_B} = \frac{\mu_0}{2\pi(r + R)} I_e I_p. \tag{57}$$

Once we use $l_B = \lambda_B/\alpha = \lambda_C/\alpha^2$, along with Equations (51) to (56), considering the limit $r + R \approx r$, the immediate consequence is that

$$F_M = \frac{\alpha}{\pi} F_E \tag{58}$$

where $F_E$ is the classic electrical force between the electron and the proton, which perfectly agrees with the prediction given in Equations (43) to (45), after considering $v_F^* = v_B^* = v_B$.

In order to validate this picture, we have looked into the relationship between the spiral cross section and the helical electrical current along with the magnetic fields produced by it. This should provide information about the magnetic flux and must agree with the well-known magnetic moment of the electron, *i. e.* the electron spin.

The helical motion of the electron must be related to a centripetal force given by

$$F_c = \frac{mc^2}{r_C} = \frac{m^2 c^3}{\hbar} = ecB \tag{59}$$

which provides the strong magnetic field

$$B = \frac{m^2 c^2}{e\hbar} = 4.413 \times 10^9 \text{ T}. \tag{60}$$

It should be noted that this centripetal force relates with the electrical force exactly as

$$F_c = \frac{1}{\alpha^3} F_E \tag{61}$$

which is extremely intense compared to the charge attraction between electron and proton (6 orders higher). However, no matter the intensity of the magnetic field and $F_c$, they have no influence on the proton, since the magnetic field is enclosed inside the toroidal solenoid predicted in Figure 4(*a*).

Even more interesting is the fact that one can use the centripetal force ($F_E = m v_B^{*2}/r_B^*$) for the orbit of the equivalent Bohr atom in Equation (61), in which $\alpha = v_B/c$ must be exchanged by the factor $\beta = v_B^*/c$. This directly provides the quantization of the angular momentum ($m v_B^* r_B^* = \hbar$) as a consequence, suggesting again that the helical motion is a universal behavior of electrons in the orbits of atoms.

Making $p$ in Figure 5(*b*) equal to zero, since $\lambda_C \operatorname{sen} \alpha \ll (\lambda_C/2\pi) \cos \alpha$, the magnetic flux per spiral of radius $r_C = \lambda_C/2\pi$ is given by

$$\Phi = BS = \frac{m^2 c^2}{e\hbar} \pi \left(\frac{\lambda_C}{2\pi}\right)^2 = \frac{h}{2e} = \Phi_0 \tag{62}$$

where $S = \pi(\lambda_C/2\pi)^2$ is the cross section of one spiral, which provides exactly one quantum flux per spiral.

Furthermore, computing the magnetic moment ($\mu_m$) for each spiral allows us to directly obtain the electron spin ($\mu_S$), as follows

$$\mu_m = IS = e \frac{c}{\lambda_C} \pi \left(\frac{\lambda_C}{2\pi}\right)^2 = \frac{e\hbar}{2m} = \mu_S. \tag{63}$$

It is interesting to observe that the motion of the electron either in the clockwise or counterclockwise directions (see red and blue arrows in Figures 4(*b*) and (*c*)), running towards the orbit, naturally provides both up and down spins for electrons, which is in agreement with what was observed in the Stern-Gerlach experiment[7,48,49]. In addition, these two electrons running the helical paths tend to cancel the total magnetic field inside the toroidal solenoid (see Figure 4(*c*) again), which must be the reason for the electron spin pairing predicted in the Pauli exclusion principle for a particular orbital[7].

On the other hand, if we look at the toroidal solenoid (Figures 4(*a*)), we can define its impedance as

$$\mathcal{L} = \frac{N\Phi_0}{I} = \frac{\Phi_0}{\alpha^2}\frac{\lambda_C}{ec} = \frac{1}{\alpha^2}\frac{h}{2e^2}\frac{h}{mc^2} = \frac{1}{\alpha^2}R_Q\frac{h}{mc^2} \qquad (64)$$

providing the period of one turn in the solenoid given by

$$\tau_B = \frac{\mathcal{L}}{R_Q} = \frac{1}{\alpha^2}\frac{h}{mc^2} \qquad (65)$$

which is the period for the electron to make one turn in the first orbit of the Bohr atom.

It should be noted that while $\tau_B$ has to do with the $l_B$ distance ($= \lambda_C/\alpha^2$) or the number $N$ of spirals ($= 1/\alpha^2$), the time constant $\tau_c$ has to do with the fundamental cyclotron frequency of the electron traveling at the speed of light in one spiral turn, which is clearly related to the time constant ($\tau_Z = \tau_c/2 = h/2mc^2$) discussed in *zitterbewegung* physics[46,50–52].

Finally, we briefly consider the distortion of the circular spirals due to the small variation in the distance between electrons and protons, *i.e.* the Bohr radius, caused by the electrical force on the spirals (see Figure 4(*a*) again). In the point nearest to the proton, the electron is under the highest electrical force than in the farthest point, in which it is under the influence of the smallest force. The small difference in the electrical force with regard to the average value ($r_B$) expected in the Bohr first orbit happen in all points of each spiral in the helical motion, providing elliptical spirals instead of circular ones. Thus, considering the semi-minor and the semi-major axes of an elliptical spiral as $b = \lambda_C/2\pi$ and $a = (1 + \alpha/2\pi)b$, respectively, the elliptical cross section ($S = \pi ab$) leads to

$$\frac{\mu_{S\,\text{elliptical}}}{\mu_S} = \frac{I\pi ab}{I\pi(\lambda_C/2\pi)^2} = \frac{a}{b} = \left(1 + \frac{\alpha}{2\pi}\right) \qquad (66)$$

or

$$\mu_{S\,\text{elliptical}} = \frac{g_e}{2} = \left(1 + \frac{\alpha}{2\pi}\right)\mu_S \qquad (67)$$

which corresponds to the first correction term of the anomalous magnetic moment predicted by Schwinger[53], where $g_e$ is the g-factor related to the electron gyromagnetic ratio[54].

*v) Electron-electron scattering, Metal-Insulator transition, and electrical resistivity $T^2$ behavior*

Considering Equations (23) and (29) together it is possible to write down

$$\rho_{ee}n = \frac{h}{2e^2}\frac{1}{Ll} = \frac{m}{\tau^* e^2}. \qquad (68)$$

Due to the SrTiO$_{3-\delta}$ to show metal-like behavior with very low charge carrier density almost temperature independent (see, for instance, figure 3(*a*) of reference 55) this compound along with conventional metals seem to be the best examples for this study.

Considering the conservation of the angular momentum, we can write

$$mv_B^* r_B^* = mv_B \frac{\lambda_B}{2\pi} = \hbar = \frac{h}{2\pi} \tag{69}$$

or

$$(\pi/3)^{1/3} mv_F \lambda = mv_B \lambda_B = mc\lambda_C = h \tag{70}$$

which means there is an equivalent Bohr radius given by $2\pi r_B^* = \lambda = 2L$ for the electrons at the Fermi level of each metal, in which the atom is seen as an electron travelling at an orbit connected to a single proton at the center of the atom. This seems to be supported by Gauss' law in which any atom can be seen as a unique proton at the center of the atom inside the volume just below the equivalent Bohr radius, *i. e.* an atom with one electron and one proton or the equivalent Bohr atom. This is supported by a description of the Metal-Insulator Transition (MIT) predicted by Mott[13] and discussed in detail by Edwards *et al.*[56] and Behnia[57].

Putting $v_F = l/\tau$, $v_B = l_B/\tau^*$, and $c = l_c/\tau^*$ into Equation (70) and remembering that $\tau = (\pi/3)^{1/3} \tau^*$, implies that

$$2Ll = 2L_B l_B = \lambda_C l_C = \frac{h\tau^*}{m} = (3/\pi)^{1/3} \frac{h\tau}{m} \tag{71}$$

which is in agreement with Equations (29) and (67). It should be noted that the smaller the equivalent Bohr radius is, the larger the mean free path is.

Such as

$$l_B = \frac{\lambda_B}{\alpha} = \frac{\lambda_C}{\alpha^2} \tag{72}$$

the length $l_c$ can be calculated based on

$$\lambda_C l_C = \frac{\lambda_B^2}{\alpha} = \frac{\lambda_C^2}{\alpha^3} \tag{73}$$

which implies

$$l_c = \frac{\lambda_C}{\alpha^3} = 6.228 \times 10^{-6} \text{ m} \tag{74}$$

that is the largest mean free path for the electron traveling at the speed of light ($v_F^* = c$) in the conductor supposedly with the highest electrical conductivity, which has charge density $n = 5.603 \times 10^{35}$ m$^{-3}$ and the maximum Fermi current of $I_F^* = 19.80$ A (ref. [46,47,58–60]).

Furthermore, the time constant between two electron-electron collisions is given by

$$\tau^* = \frac{1}{\alpha^3}\frac{\lambda_C}{c} = \frac{1}{\alpha^3}\frac{h}{mc^2} = 2.082 \times 10^{-14} \text{ s} \qquad (75)$$

or

$$\tau = (\pi/3)^{1/3}\tau^* = 2.114 \times 10^{-14} \text{ s} \qquad (76)$$

which is independent of the electrical conductor. It is noteworthy that this relaxation time allows a metal with large carrier density, *i.e.* small $L$ and many charges close to each other, to have a very large mean free path. This has to do with the collision cross section ($\sigma_S$) (ref. [38]) given by

$$\sigma_S = \frac{1}{nl}. \qquad (77)$$

After using Equations (70) and (72), one can show that

$$\sigma_S = \frac{2\alpha}{\lambda_B^2}L^4 \qquad (78)$$

leading to a typical collision radius for the electron-electron scattering which is proportional to $r_B^{*2}/r_B$.

Additionally, Equation (78) demonstrates that the larger the carrier density is, the smaller $L$ and $\sigma_S$ are, reducing electron-electron scattering, favoring electrical conduction. This is in agreement with the recent discussion by Behnia[17].

Considering Equations (71) and (73), Equation (68) can be rewritten as

$$\rho_{ee}n = \frac{h}{2e^2}\frac{1}{Ll} = \frac{h}{2e^2}\frac{2\alpha}{\lambda_B^2} = \frac{m}{\tau^* e^2} = 1.705 \times 10^{21} \text{ ohm/m}^2. \qquad (79)$$

This suggests that the mean free path of a metal has to do with the helical motion of the electron at the speed of light. Furthermore, this brings a clear insight on the mean free path of the electron. In fact, Equations (71), (73), and (79) suggest that all conducting electrons travel a mean free path proportional to $l_c$ ($l = l_c \lambda_C/\lambda$) through helical paths at the speed of light, during the time $\tau^*$ between two collisions.

Another interesting experimental aspect has to do with the MIT predicted by Mott[13], in which the electrical resistance $R$ must be equal to the quantum resistance at the transition and the critical carrier density and the equivalent Bohr radius should show a universal behavior in which $n_c^{1/3}r_B^* = $ constant (see, for instance, reference 57).

As predicted in the Equation (71), the smallest mean free path $l$ must be equal to $\lambda_C$, which is related to the longest equivalent Bohr radius $r_B^* = l_c/2\pi$ (ref. [32,33]). For this atom, the critical charge density is

$$n_c = \frac{1}{L_c^3} = \left(\frac{2}{l_c}\right)^3 = 3.312 \times 10^{16} \text{ m}^{-3}. \tag{80}$$

Thus, Equations (74) and (79) provide

$$\rho_c = \frac{h}{2e^2}\frac{2\alpha}{\lambda_B^2}\left(\frac{l_c}{2}\right)^3 = \frac{h}{2e^2}\frac{\lambda_C}{4\alpha^6} = 51{,}848 \text{ ohm.m} \tag{81}$$

which allows us to determine the electrical resistance at the MIT as

$$R_c = \rho_c \frac{l}{L^2} = \frac{h}{2e^2}\frac{2\alpha}{\lambda_B^2}(l_c/2)^3\frac{\lambda_C}{(l_c/2)^2} = \frac{h}{2e^2} = R_Q = 12{,}909 \text{ }\Omega, \tag{82}$$

which precisely agrees with the expected quantum resistance at the transition.

Furthermore, considering $l_c = 2\pi r_B^*$, it is easy to show that

$$n_c^{1/3} r_B^* = 1/\pi = 0.318 \tag{83}$$

which agrees, at least within the experimental error ($n_c^{1/3} r_B^* = 0.26 \pm 0.05$), with the result for the MIT in many metallic systems reported by Edwards et al.[56].

Finally, let us to discuss the electron-electron scattering and $T^2$ behavior of the electrical resistivity at low temperatures.

From Equation (33) it is possible to write the repulsion energy between electron-electron as

$$U = F_E L = \pi \frac{\alpha}{\lambda_C} \epsilon_F^* L \tag{84}$$

which, corrected for the volume of the one cell, provides

$$U\frac{L}{l} = \pi(\pi/3)^{2/3}\epsilon_F \frac{1}{\lambda_B}\frac{L^2}{l} = \pi(\pi/3)^{2/3}\epsilon_F \frac{1}{\lambda_B}\frac{\rho_{ee}}{R_Q} \tag{85}$$

after using Equations (23) and (68).

In fact, we noticed that $UL/l$ must be equal to the level of thermal energy per charge carrier ($N = 1$) due to the Joule effect produced by the electron-electron scattering, which allows us to write

$$U\frac{L}{l} = \epsilon_F \pi(\pi/3)^{2/3} \frac{1}{\lambda_B}\frac{\rho_{ee}}{R_Q} = C_e T = (\gamma T)T = \frac{\pi^2}{2}\frac{k_B^2}{E_F}T^2 \tag{86}$$

where $C_e = \gamma T$ is the electronic heat capacity, providing

$$\rho_{ee} = \frac{\pi^2}{2}(3/\pi)^{2/3}\lambda_B \frac{R_Q}{\pi}\frac{k_B^2}{\epsilon_F^2}T^2 = Q\frac{T^2}{T_F^2} = AT^2 \tag{87}$$

where

$$Q = \frac{\pi^2}{2}(3/\pi)^{2/3}\lambda_B \frac{R_Q}{\pi} = 6.539 \times 10^{-6} \text{ Ohm.m} \quad (88)$$

and

$$A = \frac{Q}{T_F^2} = \frac{6.539 \times 10^{-6}}{T_F^2} \text{ Ohm.m/K}^2 \quad (89)$$

or

$$A = 3.885 \times 10^{-15} \gamma^2 \text{ Ohm.m/K}^2. \quad (90)$$

Once we have found the expected behavior for the electron-electron scattering, it can be carefully compared with the experimental data for the pre-factor $A$ versus $T_F$ available in the literature. Figure 6 displays data for many conventional and non-conventional conductors taken from the several references[14–22].

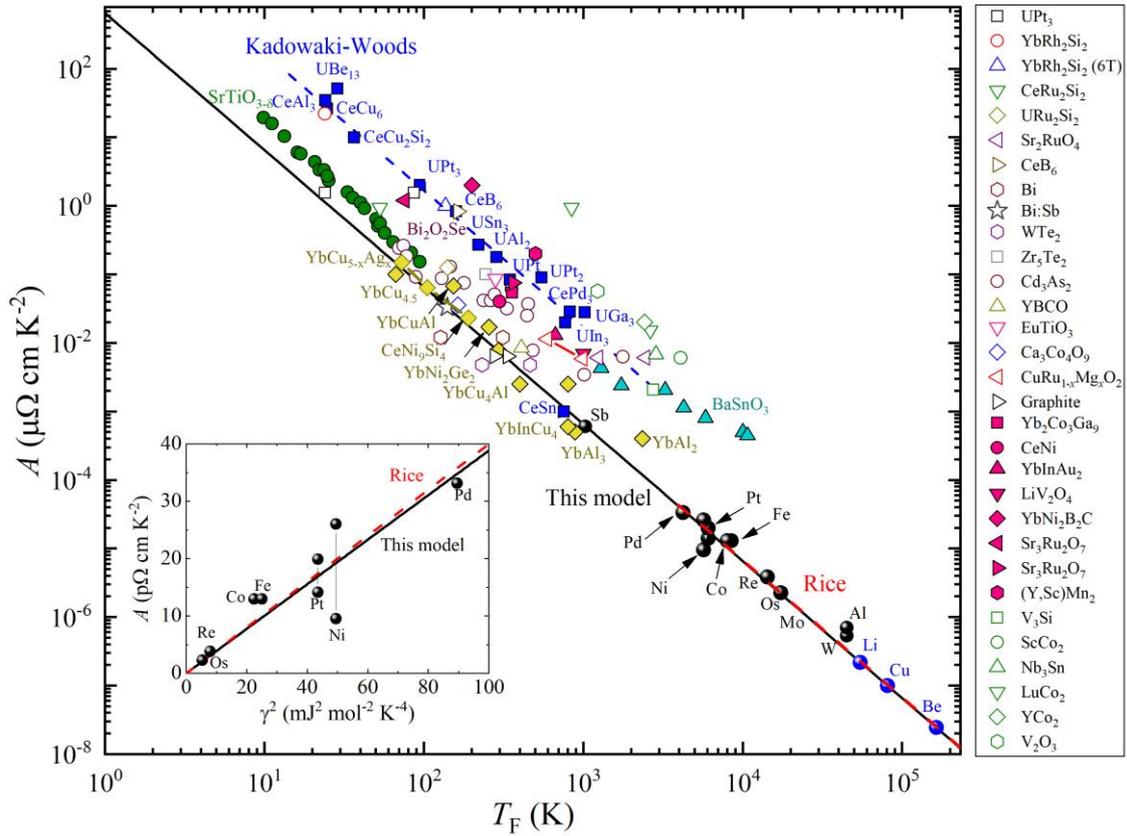

**Figure 6** – $A$ pre-factor as a function of the Fermi temperature. Black and green circle symbols are related to the classical metals and SrTiO$_{3-\delta}$, respectively (mostly reviewed by Behnia and co-authors[16–19]). Blue circle symbols are metals with very high Fermi temperature, whose $A$ pre-factor is not available experimentally. In such cases, points were calculated using Equation (89). Yellow circle symbols are due to weak magnetic correlated Yb-based compounds, reported by Tsujii *et al.*[22]. Open symbols are related to non-conventional metals, such as high temperature superconductors, strongly correlated systems, and others. Dashed red and blue lines represent the experimental observations by Rice[14] and Kadowaki-Woods[21], respectively. Black line represents the behavior given by Equation (89) and is in excellent agreement with the empirical prediction by Behnia and co-authors[16–18]. Insert compares the experimental results (dashed red line) by Rice[14] with the prediction (black line) by Equation (90).

Data for the most conducting materials (blue circle symbols), which have very high Fermi temperatures and very small $A$, were calculated using Equation (89) and $T_F$ available in the literature[5,6], and are displayed in Figure 6 only for comparison purposes. Black and green circle symbols are related to the single metallic conductors. Yellow circle symbols are due to weak magnetic correlated Yb-based compounds[22].

Dashed red and blue lines are the behaviors reported by Rice[14] and Kadowaki-Woods[21]. Black line is the prediction given by Equation (87), which fits well with the experimental data in 8 orders of magnitude of the $A$ parameter. The insert displays a comparison of the result by Rice[14] with the behavior predicted by Equation (89), which shows excellent agreement.

Open symbols are related to non-conventional conductors, in which carrier densities are temperature dependent. Due to that, they are supposedly not able to be described by the model reported in this work. The comparison of the results for the conventional metals, $SrTiO_{3-\delta}$, and Yb-based compounds with the non-conventional conductors suggests that the strong magnetic correlation, high temperature superconductivity, and other effects should play an important role on electron-electron scattering, which deviate it from the expected behavior given by the model reported here. One possibility to describe this difference is to take into account the magnetic energy ($\epsilon_M = \mu B$) in the Fermi-Dirac distribution.

Finally, the results are in excellent agreement with the recent observations by Behnia and co-authors[16–18], where pre-factor $A$ in Equation (88) can be rewritten as

$$A = \frac{\hbar}{e^2}\left(\frac{k_B}{\epsilon_F}\right)^2 \ell_{quad} \qquad (91)$$

where

$$\frac{\hbar}{e^2} = \frac{R_Q}{\pi} = 4{,}109 \text{ ohm} \qquad (92)$$

and

$$\ell_{quad} = \frac{\pi^2}{2}(3/\pi)^{2/3}\lambda_B = 1.591 \times 10^{-9} \text{ m}, \qquad (93)$$

which is the same as the empirical prediction reported in references 16 to 18.

3) **Conclusion**

In this work we report arguments to consider electrical energy as an important parameter to describe the motion of electrons near the Fermi level at zero temperature. Thermal energy has been added to the electrical energy into the Fermi-Dirac distribution, which allows us to establish equations to predict the behavior of the electrical resistivity of many metals at low temperatures.

As a consequence, Ohm's law, the Drude model, and the quantization of electrical resistance have derived naturally from quantization of the angular momentum. The description allowed to conciliate the Drude model with Fermi gas theory.

The proportionality ratio between magnetic and electrical forces predicted the helical motion of the electrons and allowed a better understanding of the mean free path. Furthermore, it was possible to show that the Joule effect is directly related to electrical and magnetic forces, which, as far as we know, has never been reported so far.

Applying the considerations to the Bohr atom, it was shown that the helical motion of the electron is related to the Sommerfeld fine structure constant. This has been confirmed by the natural appearance of the magnetic flux quantization and electron spin, as well as the anomalous behavior of the magnetic moment predicted by Schwinger.

The comparison of the model with electrical transport properties of several metallic conductors provided an excellent agreement and a better understanding of the electron-electron scattering mechanism at low temperatures. The model predicts the behavior of pre-factor $A$ of the quadratic temperature electrical resistivity dependence at low temperatures, due to electron-electron scattering in conductors, in a wide range (8 orders of magnitude). The model is in excellent agreement with previous results by Rice and with the recent observations by Behnia and co-authors.

This work opens a new scenario for the description of electron motion, as well as introduces new insights into the quantum mechanics and relativistic behavior of the electrons. The dependence of the model with the Sommerfeld fine structure constant suggests that this work must be of interest for several areas, such as solid-state physics, particle physics, and quantum electrodynamic theory.

**Acknowledgements**

The authors are thankful to J. J. Neumeier for many valuable comments. L. M. S. Alves is post-doc at EEL-USP (Proc. 21.1.419.88.8). F. S. Oliveira is post-doc at UNICAMP (FAPESP 2021/03298-7). M. S. da Luz is CNPq fellow (Proc. 311394/2021-3).